\documentclass[12pt,preprint]{aastex}

\shorttitle{The Centaurus Group and NGC 5128}
\shortauthors{Woodley}

\received{2006 June 20}
\begin{document}

\title{The Centaurus Group and the Outer Halo of NGC 5128:  Are they Dynamically Connected?}

\author{Kristin A.~Woodley}
\affil{Department of Physics \& Astronomy, McMaster University,
  Hamilton ON L8S 4M1}
\email{woodleka@physics.mcmaster.ca}

\begin{abstract}
NGC 5128, a giant elliptical galaxy only $\sim 4$ Mpc away, is the
dominant member of a galaxy group of over 80 probable members.  The
Centaurus group provides an excellent sample for a kinematic comparison between the
halo of NGC 5128 and its surrounding satellite galaxies.   A new
study, presented here, shows no kinematic difference in 
rotation amplitude, rotation axis, and velocity dispersion between
the halo of NGC 5128, determined from over $\sim340$ of its globular
clusters, and those of the Centaurus group as a whole.
These results suggest NGC 5128 could be behaving in part as the inner component to
the galaxy group, and could have begun as a large initial
seed galaxy, gradually built up by minor mergers and satellite
accretions, consistent with simple cold dark matter models.   The mass
and mass-to-light ratios in the B-band, corrected for 
projection effects, are determined to be $(1.3\pm0.5) \times 10^{12}$
M$_{\sun}$ and $52\pm22$ M$_{\sun}$/L$_{\sun}$ for NGC 5128 out to a
galactocentric radius of 45 kpc, and
$(9.2\pm3.0) \times 10^{12}$ M$_{\sun}$ and $153\pm50$
M$_{\sun}$/L$_{\sun}$ for the Centaurus group, consistent with
previous studies.
\end{abstract}


\section{Introduction}

The study of galaxy groups can provide essential insight into galaxy
formation, particularly because many groups are much less dynamically
evolved than rich clusters and thus are closer to their earliest
conditions.  The density of formation
environment affects the dynamical evolution of galaxies; for example, 
the luminosity and color distributions of galaxies differ in
field versus group environments \citep{girardi03}.  Furthermore, for
groups with large central members, the properties of the
central galaxy correlate with 
properties of their satellite galaxies, such as the
satellite-type fraction found in the group \citep{weinmann06}, the alignment of satellite galaxies
preferentially along the dominant galaxy's
major axis \citep{yang06,brainerd05,zentner05}, and the rotation of
satellite galaxies in the same direction as the host halo \citep{warnick05}.  

The connections between the centrally
dominant galaxy in a group environment and its surrounding satellite 
galaxies raise the question whether there might be a more direct kinematic 
connection between them as well.  Kinematic similarities between the
group and its brightest group member would be consistent with both cold dark matter
models, suggesting the central giant and its satellite group are
part of the same large-scale structure,
and the hierarchical formation scenario, whereby the central giant
would build up by accretion of satellites in the group.  

NGC 5128 is an E/S0 giant galaxy in the positional and dynamical center
of the Centaurus group of galaxies.  Its remarkably close proximity
(just 4 Mpc distant) also provides a genuinely rare opportunity to 
study and compare the dynamics
of the halo of NGC 5128 with the
dynamics of the surrounding group at a level of detail that is
difficult elsewhere.   

\begin{figure}
\plotone{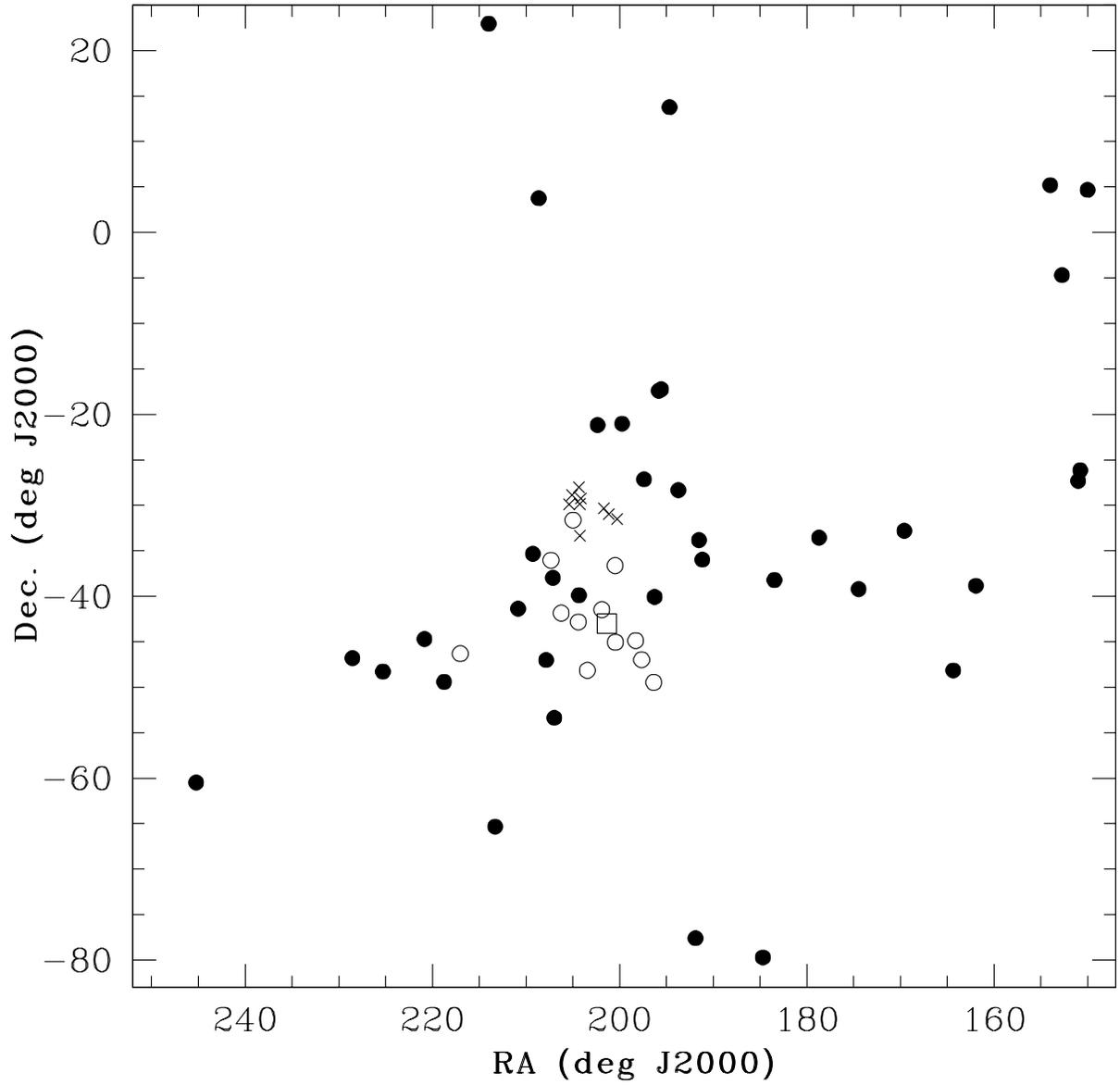}
\caption{The positions of galaxies surrounding NGC 5128 with available
radial velocity measurements.  NGC 5128 is the {\it open square}, the
confirmed galaxies in the Centaurus group are {\it open circles}, the
confirmed members in the M83 group are {\it crosses}, and the
surrounding galaxies that are not confirmed to belong to either the
Centaurus or M83 groups are {\it solid circles}.}
\label{fig:galpos}
\end{figure}

The Centaurus group consists of 25 confirmed galaxy members \citep{kar06}, 13 of which
have radial velocity measurements used in this kinematic study; see Figure~\ref{fig:galpos}.
Within a projected distance of $\sim1$ Mpc from NGC 5128 lies NGC 5236
(M83).  NGC 5236 is a dominant spiral galaxy surrounded by 9 other confirmed members within the M83
group.  Surrounding these two systems are 53 smaller
galaxies that do not have confirmed membership in either the Centaurus
group or the M83 group.  Therefore the bounds separating these two groups are
still quite uncertain, and it has been suggested that these two
systems create a dumbbell system, similar to our own Milky Way and M31
\citep{kar06}. 

Only one other kinematic comparison between the halo of a dominant galaxy and its
surrounding satellite galaxies (M87 in the Virgo cluster) has been
presented in the literature \citep{cote01}.  Virgo is a much more
dynamically evolved environment than the Centaurus group.  The Centaurus
group, therefore, provides the chance to carry the comparison a step further than
just alignments of satellites and simple number statistics by looking
at an environment closer to its initial structure.  

\section{Kinematics}
\label{kin}

NGC 5128 has many hundreds of 
globular clusters (GCs) that can be used to define 
its halo kinematics and dynamics.  The most
recent studies \citep[M. Beasley et al. 2006, in preparation;][]{whh05,pff04I} 
now raise the confirmed globular cluster (GC) membership to $\sim340$
members with known radial velocities, which make up the sample used in
this paper (this list includes $\sim100$ new clusters to be
published in M. Beasley et al. (2006, in preparation), written in tandem
to this study).  A full catalogue of all GCs in NGC 5128, along with
radial velocities and photometry will be published in
K. Woodley (2006, in preparation).  
The Centaurus group and neighbouring M83 group,
have many smaller galaxies with radial velocity
measurements.  This combined material enables a kinematic and dynamic 
comparison between the brightest group member and its satellite group.

The kinematics of both the globular cluster system (GCS) of
NGC 5128 and the Centaurus group of galaxies are defined using the normal condition
\begin{equation}
\label{eqn:kin}
v_r(\Theta) = v_{sys} + \Omega R  sin(\Theta - \Theta_o)
\end{equation}
\citep[e.g.][]{richtler04,cote01}
where $v_r$ is the observed radial velocity of the object in the
system, $v_{sys}$ is the systemic velocity, R is the
projected radial distance of each object from the center of the
system, and $\Theta$ is the projected azimuthal angle of the object measured
in degrees East of North.  The quantities obtained by applying this
equation to the system, with the known $v_{sys}=541$ km s$^{-1}$ for
NGC 5128 \citep{hui95} held constant, are
$\Theta_o$, the rotation axis of the system measured, and the 
product $\Omega R$, the rotation amplitude of the objects in the system.

The projected velocity dispersion is
calculated using the maximum likelihood dispersion estimator described
by \cite{pryor93},
\begin{equation}
\label{eqn:veldisp}
\sum_{i=1}^{N} \frac{(v_i - v_{sys})^2}{(\sigma_{v}^2 + \sigma_{v_{r_i}}^2)^2} = \sum_{i=1}^{N} \frac{1}{(\sigma_{v}^2 + \sigma_{v_{r_i}}^2)}
\end{equation}
where $N$ is the number of clusters in the sample, $v_i$ is the
cluster's radial velocity {\it after subtraction of the rotational
component}, and $\sigma_{v_{r_{i}}}$ is the uncertainty in the velocity
measurement.  $\sigma_v$ is the projected velocity dispersion
determined by iterating through values of 
$\sigma_v$ to find the equality of Equation~\ref{eqn:veldisp}.  The
systemic velocity is again held constant at 541 km s$^{-1}$.  The
uncertainties were calculated from the variance of the dispersion 
using Equations~6, 8, 9, \& 10 from \cite{pryor93}.

\subsection{The Globular Cluster System of NGC 5128}
\label{kinGCS}

The kinematic condition of Equation~\ref{eqn:kin} was applied to the
GCS of NGC 5128.  A total of 343 GCs
with radial velocity membership in NGC 5128 have been used in the 
kinematic determination.  The radial
velocities used are the weighted averages from all
previous measurements studies of NGC 5128 and are described completely
in K. Woodley et al. (2006, in preparation).
The GCs were assigned weights in the kinematic fitting by
considering two components to their associated uncertainties: the
individual observational uncertainty in $v_r$, and the random velocity
component of the GCS evident by the large dispersion in the confirmed 
GC velocities.  In almost all cases, this latter term dominates,
leaving the clusters with very similar weights.

The fitted values of $\Omega R$ and $\Theta_o$ were determined for 
projected galactocentric radial bins with inner and outer radii listed in
column~1 of Table~\ref{tab:GC}.  Successive columns give the mean
radius and the outer radius in kpc assuming a distance of 3.9 Mpc, the number of GCs, 
the rotation amplitude in km s$^{-1}$,
the rotation axis of the GCs in each circular radial
bin in degrees East of North, the velocity dispersion in km s$^{-1}$,
the projected mass correction factor, the pressure supported component of mass, the 
rotationally supported component of mass, and the total mass 
(see Section~\ref{mass} for the mass discussion). 

\begin{figure}
\begin{center}
\epsscale{.8}
\plotone{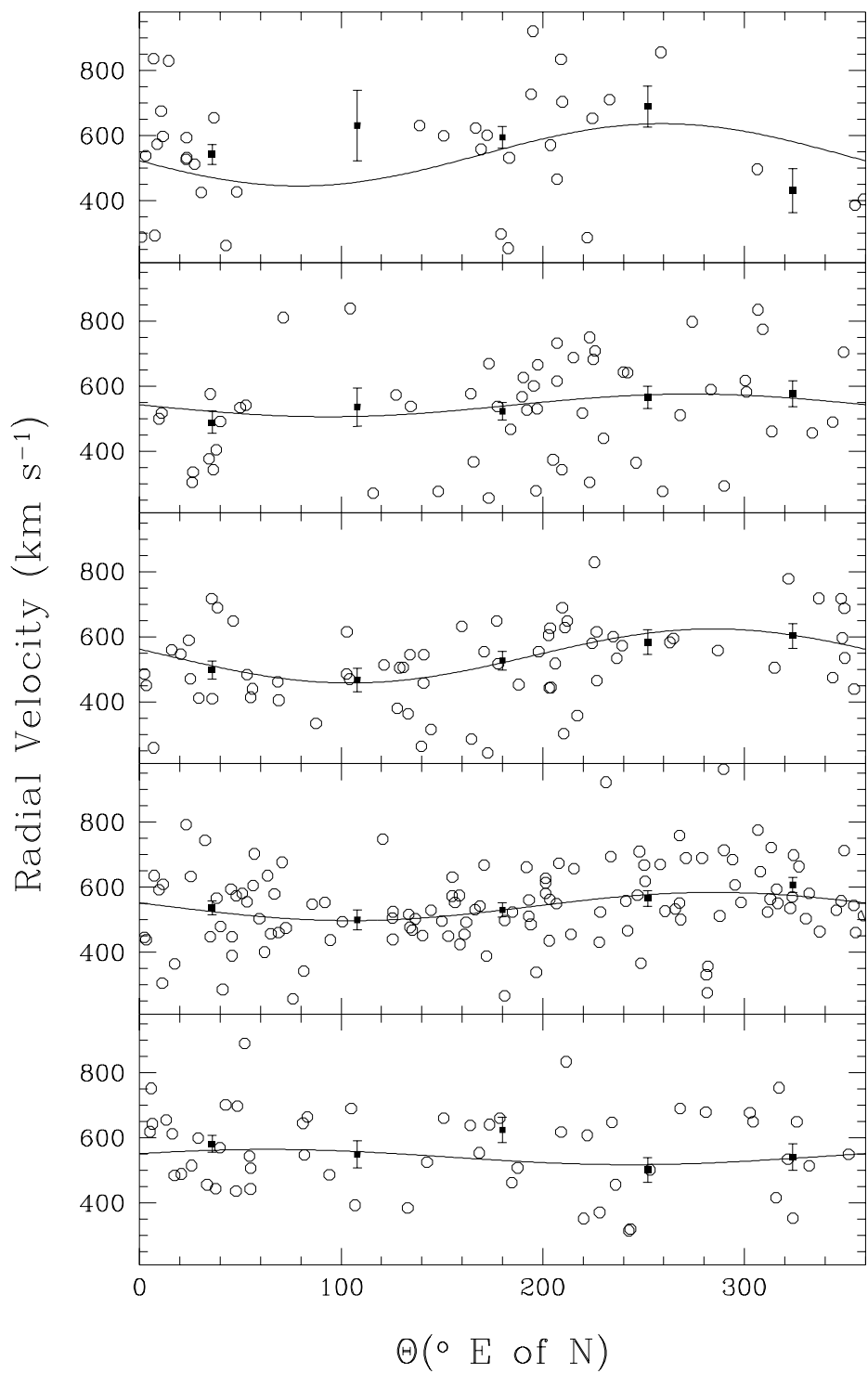}
\caption{The sine curve fit for the GCs in NGC 5128 indicated by 
the {\it open circles}, binned radially with fixed systemic velocity.  
The radial bins are ({\it bottom panel}) 0-5 kpc, ({\it 2$^{nd}$
  panel}) 5-10 kpc, ({\it 3$^{rd}$ panel}) 10-15 kpc, 
({\it 4$^{th}$ panel}) 15-25 kpc, and ({\it top panel}) 25-50 kpc.  
The {\it solid squares} are the weighted velocities in 72$^o$ bins.}
\label{fig:kinGC}
\end{center}
\end{figure}

The radial bin 0-5 kpc has a different
rotation solution from bins 5-10, 10-15, 15-25, 25-50, and 0-50 kpc.
In bin 0-5 kpc, these GCs nominally rotate
around a similar rotation axis as the remaining
bins, but in an opposing direction.
This result may be real for this innermost radial 
bin since it contains 55 clusters distributed evenly in angle over that portion
of the sky (see Figure~\ref{fig:kinGC}).  However, the rotation
amplitude is scarcely different from zero, and the system is clearly
dominated by internal random motion there.

Fig.~\ref{fig:kinGC} displays the sine curve fits defined by Equation~\ref{eqn:kin}.  
Beyond a galactocentric distance of 15
kpc, the observational bias of the detections (most of the known outer
clusters lie along the isophotal major axis of the galaxy, which is
where previous searches have mainly concentrated) is evident by a clear
lack of known GCs near the photometric minor axis of 
$\Theta = 119$ and $299\pm5^o$ \citep{dufour79}.  This bias leads to higher uncertainties
in the determined kinematic quantities in these outer regions.
Nevertheless, it is notable that the
kinematics of the metal-poor and metal-rich populations of GCs 
in NGC 5128 do not show any extreme differences in the
outer-halo region and I treat them as a combined sample in the
following discussion.  Full details of the kinematic solutions of the metallicity 
sub-populations are discussed in K. Woodley et al. (2006, in preparation).

The velocity dispersion of the GCS ranges from
105 to 166 km s$^{-1}$, and the entire system as a whole has a mean dispersion of
$123\pm6$ km s$^{-1}$.  These results show striking resemblances to
those of \cite{pff04III}: from an earlier
and smaller sample, they find the 
velocity dispersion of the GCs
in NGC 5128 to range from $100-150$ km s$^{-1}$ out
to a projected radius of 50 kpc, with the largest dispersion at 20 kpc
(see their Figure~20).

\subsection{The Satellite Galaxies}
\label{kinGG}

The same kinematic analysis was applied to the Centaurus group of
galaxies.  NGC 5128, as the brightest member of the Centaurus group,
was considered the center of mass of the system, and the surrounding
small galaxies as its satellites.

The distances of the surrounding galaxies were taken from Table~2 of
\cite{kar06}, while the distance to NGC 5128 itself, was assumed to be
3.9 Mpc from \cite{rej04}.  The positions and radial velocities
of the galaxies were taken from Table~1 of \cite{kar04}.  
The rotation amplitude
and rotation axis were determined for the confirmed galaxies 
with known radial velocities that are considered Centaurus group
members; \cite{kar06} have confirmed 13
members in the Centaurus group and 8 members in the M83 group with
published radial velocities in \cite{kar04}). 

Kinematic solutions were also determined
for all satellite galaxies around NGC 5128, not just those with
confirmed membership in \cite{kar06}, in two different
ways.  The first method has grouped the surrounding galaxies into two
independent bins.  Bin 1 contains the nearest 27 galaxies in projected
radii to NGC 5128, excluding any confirmed members of the M83
group.  Bin 2 contains the $28^{th}$ to $53^{rd}$ nearest galaxies to
NGC 5128, again excluding any confirmed members of the M83 group.  The
results for the Centaurus group, M83 group, and the two independently
binned samples are listed in Table~\ref{tab:IGG}.  In all subsequent
plots, the radial values plotted for the GCs, M83 group, Centaurus
group, Bin 1, and Bin 2 are the mean positions of radii in the sample
for the rotation amplitude, rotation axis, and velocity dispersion,
while the data is plotted at the outermost radii in
each bin for the mass plot.

The second method
determines the same kinematics for the galaxies surrounding NGC 5128, 
beginning with the first 14
galaxies in projected radial distance.  Successively
adding one galaxy radially outward from NGC 5128 to the first 14 
galaxies and recalculating the kinematic 
quantities was performed out to a maximum of 62 galaxies, {\it
  including M83 group members}, with
results as shown in Table~\ref{tab:GG} (although only every third point is
listed in Table~\ref{tab:GG}, all points are plotted in
subsequent figures)\footnote{The dwarf galaxy, Tucana was not 
included in the present analysis because of its large projected distance 
from NGC 5128.}.  In all subsequent plots, the radial
values plotted for the cumulative galaxy bins are the radial value of
the last galaxy in the sample.  Tables~\ref{tab:IGG} \&~\ref{tab:GG}
list, in successive columns, the bin description in
Table~\ref{tab:IGG} or the number of galaxies included in
the calculation in Table~\ref{tab:GG}, the
mean projected radius and outer projected radius in Table~\ref{tab:IGG} and the outer projected
radius in Table~\ref{tab:GG} of the sample in kpc,  
the rotation amplitude in km s$^{-1}$, the rotation axis in 
degrees East of North, the velocity dispersion in km s$^{-1}$, the
volume density slope, the pressure supported mass, the rotionally supported
mass, the total mass, the B-band
luminosity in solar luminosity units, and mass-to-light ratio.  
There appears to be no significant kinematic
difference within measurement uncertainty for all the subgroups of galaxies used
in this study. 

\subsection{Kinematic Discussion}
\label{kin_disc}
\begin{figure}
\plotone{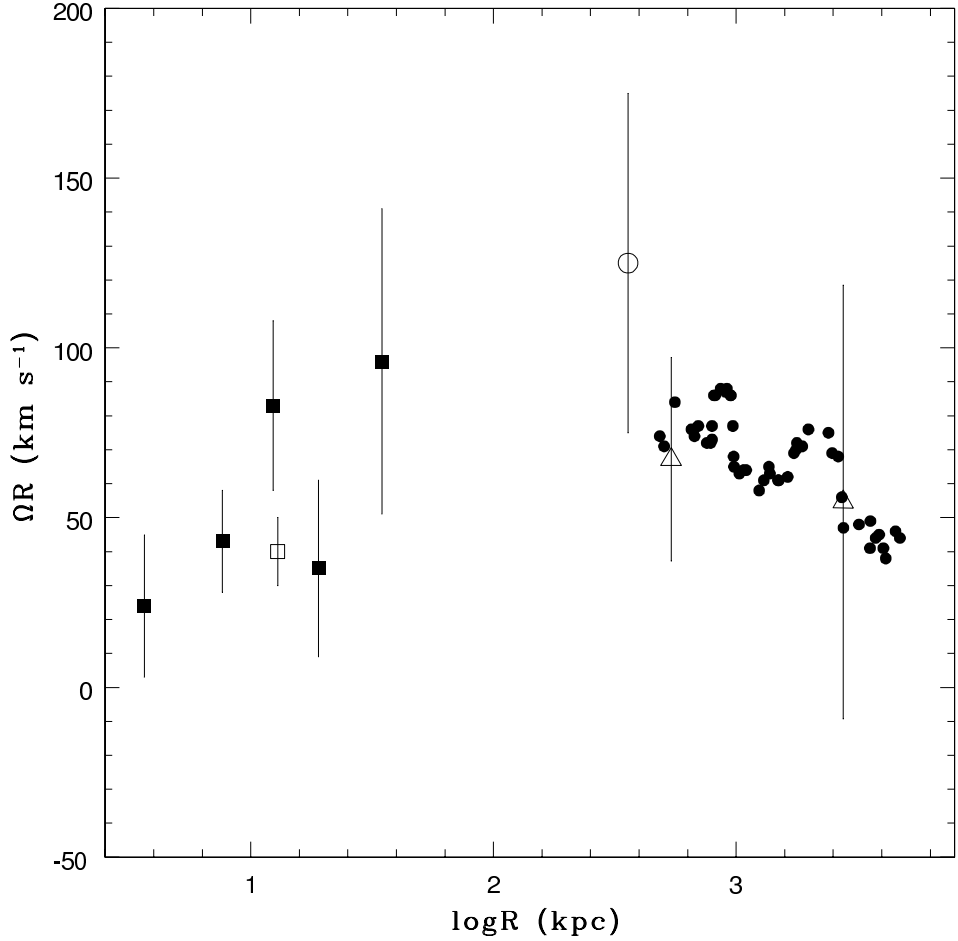}
\epsscale{1}
\caption{Rotation amplitude, $\Omega R$, as a function of projected radius from the
  center of NGC 5128 plotted for the 343 GCs in NGC 5128 
  ({\it open square}), and in independent radial bins ({\it closed squares}), 
  see Table~\ref{tab:GC}).
  Also plotted are the confirmed galaxy members from the
  Centaurus group in \cite{kar02} ({\it open circle}), with the
  cumulative addition of other probable satellite galaxies in 
  angular distance from NGC 5128 ({\it closed
  circles}), including the M83 group, see Table~\ref{tab:GG}.  The
  galaxies are also independently binned for the nearest 27 galaxies
  from NGC 5128 in projected radius (Bin 1), and then the following 26 galaxies
  in the next bin (Bin 2) ({\it open triangles}), excluding the M83
  group.  The mean rotation amplitude is
  67$\pm$27 km s$^{-1}$ for the cumulative galaxy points.  
  The radial values plotted are the average of
  the objects in each bin, with the exception of the additive
  galaxy points ({\it solid circles}), which are plotted at the radial
  position of the last galaxy in each bin.} 
\label{fig:rotamp}
\end{figure}

\begin{figure}
\plotone{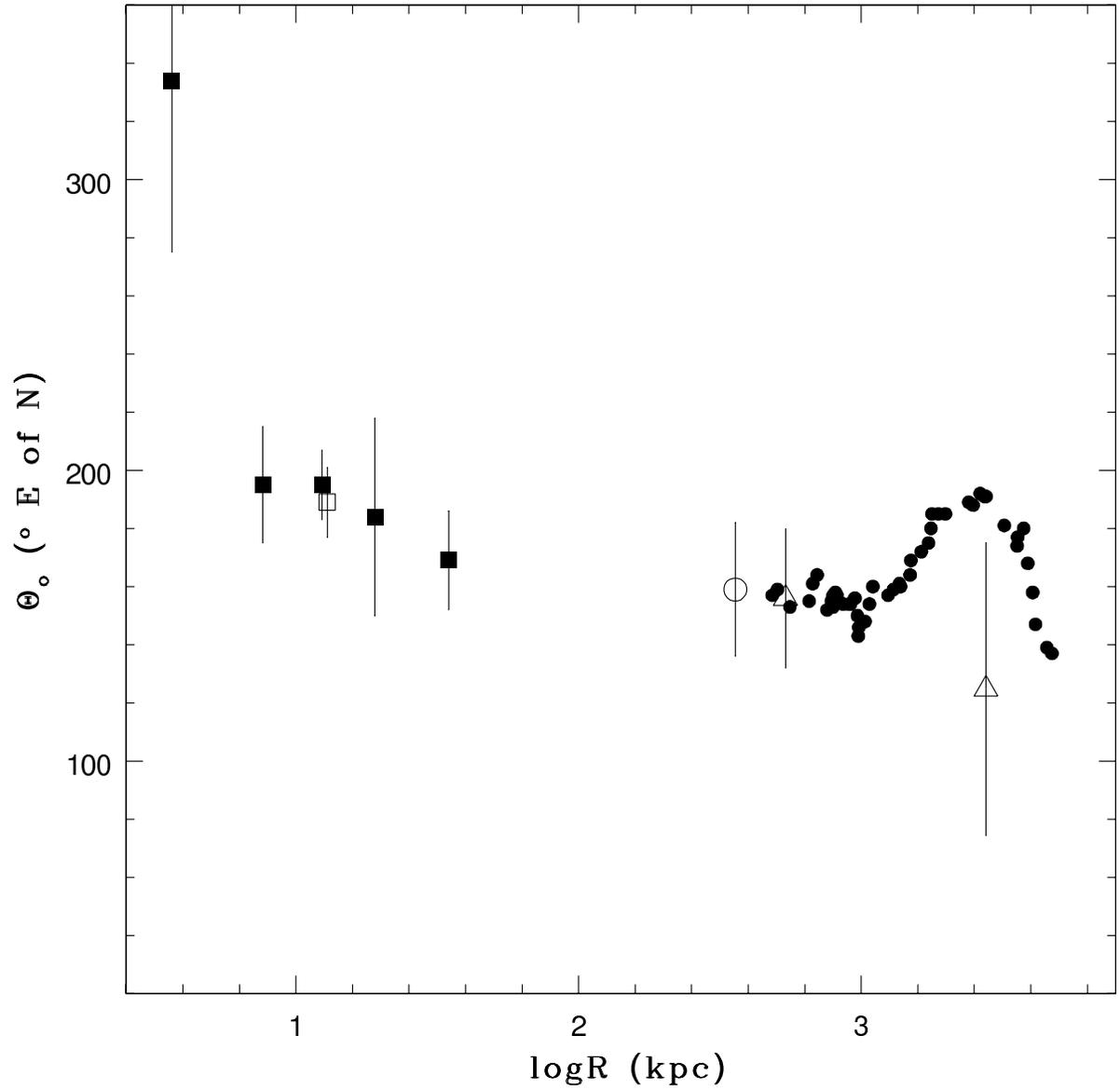}
\caption{Rotation axis, $\Theta_o$, measured in degrees East of North, as a
  function of projected radius from the
  center of NGC 5128.  Symbols are the same as in
  Fig.~\ref{fig:rotamp}.  The mean rotation axis is
  163$\pm$22$^o$ E of N for the cumulative galaxy points. 
  The radial values plotted are the average of
  the objects in the each bin, with the exception of the additive
  galaxy points ({\it solid circles}), which are plotted at the radial
  position of the last galaxy in each bin.}
\label{fig:rotaxis}
\end{figure}

Figure~\ref{fig:rotamp} shows the rotation amplitude as a function of
projected radius from the center of NGC 5128 for the GCs in
NGC 5128, as well as the Centaurus group, Bin 1, Bin 2, and the
cumulative galaxy bins including up to 62 galaxies surrounding NGC 5128.  
Figure~\ref{fig:rotaxis} shows the rotation axis as a function of
projected radius from NGC 5128. 

The kinematic solution of the outer halo of NGC 5128, determined by
its GC population, closely matches that of the satellite group.  
This suggests that the entire Centaurus group of galaxies
could be thought of as an
outward extension of its massive central galaxy; or conversely, that the halo
of NGC 5128 behaves like an inward extension of the entire group.
This result would be consistent with the idea that NGC 5128, starting
from an initial (perhaps large) ''seed'' galaxy, could
have built up from the accretions of many nearby small satellites that would
have preferentially come in along the major axis of the group.
Alternatively, if NGC 5128 formed from the merger of a {\it small} number of
{\it large} galaxies (perhaps only 2 or 3 galaxies), it would not as
easily explain the kinematic similarities since these few would be
less likely to come together along this same axis.

\begin{figure}
\epsscale{.9}
\plotone{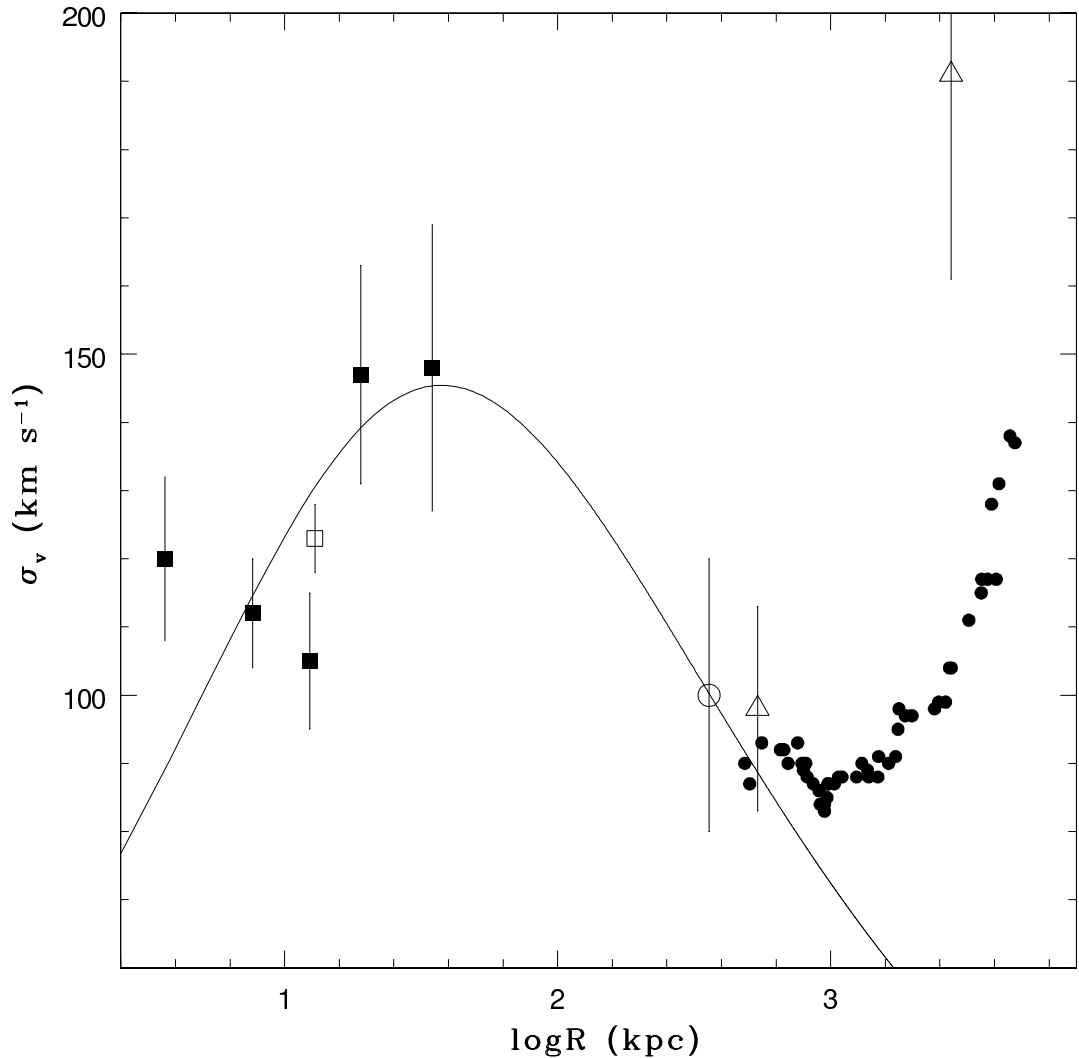}
\epsscale{1}
\caption{\small{The velocity dispersion, $\sigma_v$, as a function of radius from the
  center of NGC 5128.  The  symbols are the same as in
  Fig.~\ref{fig:rotamp}.  The average velocity
  dispersion of all points is $103\pm14$ km s$^{-1}$.  The solid line
  is the dark matter halo fit using an NFW model with a scale radius
  of 14 kpc.  The outermost galaxies are not believed to be virialized
  (see Section~\ref{sub:mass_cen} for details).  The radial values plotted are the average of
  the objects in each bin, with the exception of the additive
  galaxy points ({\it solid circles}), which are plotted at the last galaxy's radial position.} }
\label{fig:veldisp}
\end{figure}

Figure~\ref{fig:veldisp} shows the velocity dispersion as a function
of projected radius from the center of NGC 5128, plotted for the same
objects as Figs.~\ref{fig:rotamp} \& \ref{fig:rotaxis}.  The
cumulatively binned
galaxies exhibit a similar range of velocity dispersion to the GCs 
in NGC 5128, from $83-138$ km s$^{-1}$.  Yet, the
independently binned outer galaxies (Bin 2) have a much higher
dispersion then the inner galaxies in Bin 1.  This increase in
velocity dispersion in the outer region is hinted at by the outer
cumulative galaxy bins which exhibits a sharp upturn beginning near
1400 kpc.  This sharp change
in velocity dispersion, and subsequently, the mass (see
Section~\ref{mass}), does not coincide with the addition of galaxies
in the M83 group, which begins in the cumulative galaxy bin 18 
or with the large spiral, NGC 5236, is added in the cumulative galaxy bin 27 
plotted at the radial value of 907 kpc.  The upturn in velocity
dispersion is likely to be due to the non-virialization of the outer
galaxies (refer to the discussion in Section~\ref{sub:mass_cen}).  

The solid line in Fig.~\ref{fig:veldisp} represents a standard dark matter halo
model fit to
the data (Navarro, Frenk, \& White 1997, hereafter NFW), nicely
tracing the dispersion of
the halo of NGC 5128 and the {\it inner} galaxies of the Centaurus
group as well, for a scale radius, $r_s = 14$ kpc.  
However, the entire Centaurus group over all radii and halo of NGC 5128 cannot be fit by
a single NFW curve.   That is, the group as a whole likely has a density
profile that is shallower than $R^{-3}$ at large radius.

The only
other direct kinematic comparison between the GCS
of a dominant galaxy with its surrounding satellite galaxies is for M87 in the Virgo
cluster by \cite{cote01}.  \cite{cote01} showed the velocity
dispersions of the halo GCs in M87 and the satellite
galaxies clearly matched their constructed 2-component mass model for the Virgo
cluster.  But, unlike our results for NGC 5128 and the Centaurus group,
the Virgo galaxies have a dispersion that is much higher than the GCs in the M87 halo.
These results are not surprising because M87 and the Virgo cluster
contain a large X-ray halo, tracing the large dark matter component in
the Virgo cluster that dominates the potential well beyond the halo
itself.  In the Centaurus group, we clearly do not see this same
effect of a very massive, extended outer component to the same degree.

\section{Mass Determination}
\label{mass}
The total mass, $M_t$ of NGC 5128 and the Centaurus group can be determined by the addition of the
mass components supported by rotation, $M_r$, and random internal motion
(''pressure''), $M_p$,
\begin{equation}
\label{eqn:total_mass}
M_{t} = M_{p} + M_{r}.
\end{equation}

The mass supported by pressure was determined here by the Tracer Mass
Estimator, developed by \cite{evans03} as a generalized projected mass
estimator that determines the mass enclosed within the outermost
object in the sample.  The (spherically symmetric) tracer population, in the Tracer
Mass estimator, does not necessarily follow the same mass profile as the system.  Although the
Tracer Mass Estimator is not the typically selected mass estimator to
determine galaxy group mass, it allows a direct comparison with the masses
determined from the globular clusters.  The pressure
supported mass in the Tracer Mass Estimator formulation is
\begin{equation}
\label{eqn:tme}
M_{p} = \frac{C}{GN} \sum_{i}(v_{f_i} - v_{sys})^2R_i
\end{equation}
where $N$ is the number of objects in the sample and $v_{f_i}$ is the
radial velocity of the tracer object with the rotation
calculated in Section~\ref{kin} removed.  
For an isotropic population of
tracer objects, assumed in this study, the value of C is
\begin{equation}
\label{eqn:C}
C = \frac{4(\alpha + \gamma)(4 - \alpha -\gamma)(1-(\frac{r_{in}}{r_{out}})^{(3-\gamma)})}{\pi(3-\gamma)(1-(\frac{r_{in}}{r_{out}})^{(4-\alpha-\gamma)})}
\end{equation}
where $r_{in}$ and $r_{out}$ are the 3-dimensional radii corresponding
to the 2-dimensional projected radii ($R_{in}$ and $R_{out}$) of the innermost and
outermost tracers in the sample.  The value of $\gamma$ is
defined as the slope of the volume density distribution,
\begin{equation}
\label{gamma}
\rho(r) \propto (\frac{1}{r})^\gamma,  r_{in} \lesssim r \lesssim r_{out}.
\end{equation}
$\gamma$ is found by determining the surface density slope of the sample and
deprojecting the slope to three-dimensions in order to obtain the volume density slope.  
In Equation~\ref{eqn:C}, $\alpha$ describes the potential of the
system, and is set to zero in this study representing an isothermal 
halo potential in which the system has a flat
rotation curve at large distances.  Assuming isotropy in an anisotropic
system can, however, overestimate or underestimate the mass by $30\%$
for plausible ranges in anisotropy \citep{evans03}.  

Another contributor to the
uncertainty in the pressure supported mass is from the values assumed for
$r_{in}$ and $r_{out}$.  \cite{evans03} suggest that $r_{in}\simeq
R_{in}$ and $r_{out}\simeq R_{out}$ for distributions taken over a
wide angle (i.e. covering the outer halo reasonably well).  However, these approximations when taken
at intermediate radial ranges within the distribution lead to an
underestimate of the determined mass since the true $r_{out}$ can be
quite a bit larger than $R_{out}$.
I have estimated the necessary correction factors for the values of $r_{in}$ and
$r_{out}$ by generating Monte Carlo distributions of both the GCS 
of NGC 5128, and the group environment.  For the 
GCS of NGC 5128, 340 globulars
were randomly placed in a spherically symmetric system extending out
to 50 kpc with a $r^{-2}$ projected density.  The distributions were binned in 0-5, 5-10, 10-15,
15-25, 25-50, and 0-50 kpc.  For the satellite group of
galaxies, a similar spherically symmetric
distribution of galaxies was
generated for each subgroup in Table~\ref{tab:GG}.  The smallest and largest
3-dimensional and projected radii were determined in each radial bin and used
to calculate the value of C in Equation~\ref{eqn:C} for the real
positions and the projected positions.  I used a total of 500 simulated
systems.  The correction factors, $M_{corr}$, are multiplied onto all
the pressure supported component masses in this study, and are shown
in Table~\ref{tab:GC} for
the GCs.  The correction factors for the galaxies are quite 
small, as expected, falling between values of 1 and 1.3.

A recent study by \cite{yencho06} shows the use of mass estimators,
such as the Tracer Mass Estimator, can lead to incorrect mass
estimates of up to $\sim20\%$ by neglecting substructure within the
system in question.  For the sample
sizes used in this study, any such
uncertainties are less than the uncertainties claimed for anisotropic
orbital motion and the statistical uncertainty in $M_p$ itself.

The mass component supported by rotation was calculated from the rotational
component of the Jeans Equation,
\begin{equation}
\label{eqn:rje}
M_{r} = \frac{R_{out}v^{2}_{max}}{G}
\end{equation}
where $R_{out}$ is the outermost tracer radius in the sample and
$v_{max}$ is the rotation amplitude, calculated in Section~\ref{kin}.
These $M_r$ values are listed in Tables~\ref{tab:IGG} \&~\ref{tab:GG}.

\subsection{The Globular Cluster System of NGC 5128}
\label{mass:GCS}

\begin{figure}
\plotone{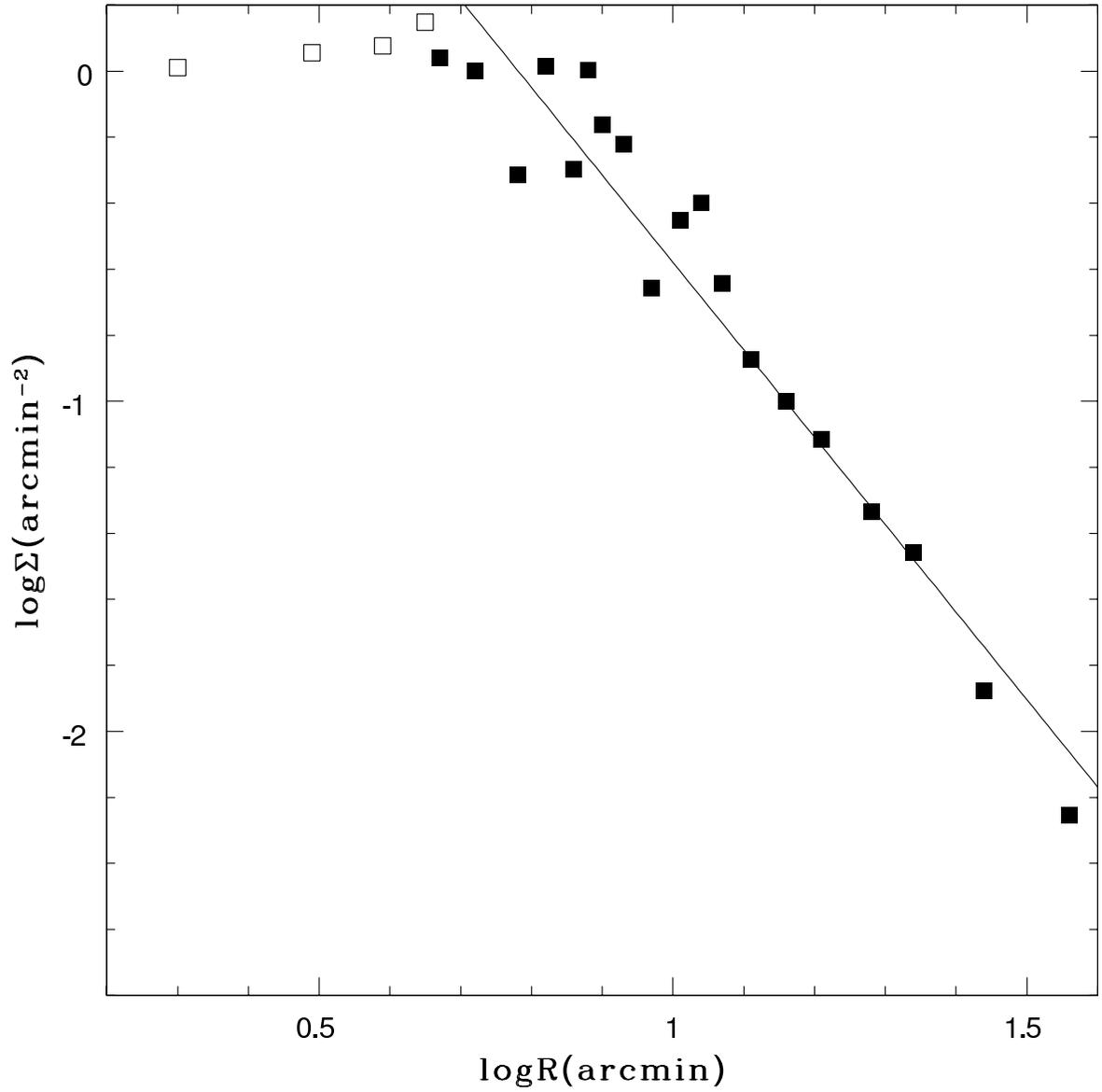}
\caption{The surface density, $\Sigma$, of the GCs in NGC 5128 
as a function of galactocentric radius.  Each bin is equally weighted
and fit to a power law with slope $-2.65\pm0.17$.  Data less than 5
kpc from the galaxy center ({\it  open circles}) has been ignored for the fit.}
\label{fig:surden}
\end{figure}

The mass of NGC 5128 follows from its tracer population of
GCs by determining the surface
density of the GCS and deprojecting the slope from
2-dimensions to 3-dimensions.
The surface density distribution of the entire GCS was determined by
binning the known clusters into circular annuli of equally populated bins, giving each bin
the same statistical weight to minimize biases \citep{maiz05},
although, as noted above,
the distribution may still have {\it spatial} biases in favor of the
major axis at large radii.  The profile is shown in
Figure~\ref{fig:surden}.  In the innermost region, within 5 kpc, 
there is incompleteness due to the
obscuration of the dust lane and this region is therefore excluded from the mass determination.  
The surface density fits well to a power
law outside of 5 kpc with a slope of $-2.65\pm0.17$ \citep[reduced
$\chi^2=0.04$ from a Marquardt-Levenberg fitting routine from][]{press92}.  A value
of $\gamma=3.65$, was then used in the mass calculations.

The rotationally supported mass was calculated from
Equation~\ref{eqn:rje}, and the determined rotation amplitude in
each bin shown in Table~\ref{tab:GC}.  The determined masses are
actually lower limits as the inclination angle of the true rotation
axis with the plane of the sky is not known.   
\cite{pff04II} argue, using planetary nebula data, that NGC
5128 is triaxial in nature and is viewed nearly edge-on.  If so, 
the rotation measured would be a good estimate of the true rotation of the
system.  In any case, $M_r$ is small compared with the pressure
component, $M_p$.

The total mass of the system determined within the outermost cluster in
each bin is listed in Table~\ref{tab:GC}.

\subsection{The Centaurus Group Mass}
\label{sub:mass_cen}

\begin{figure}
\epsscale{.9}
\plotone{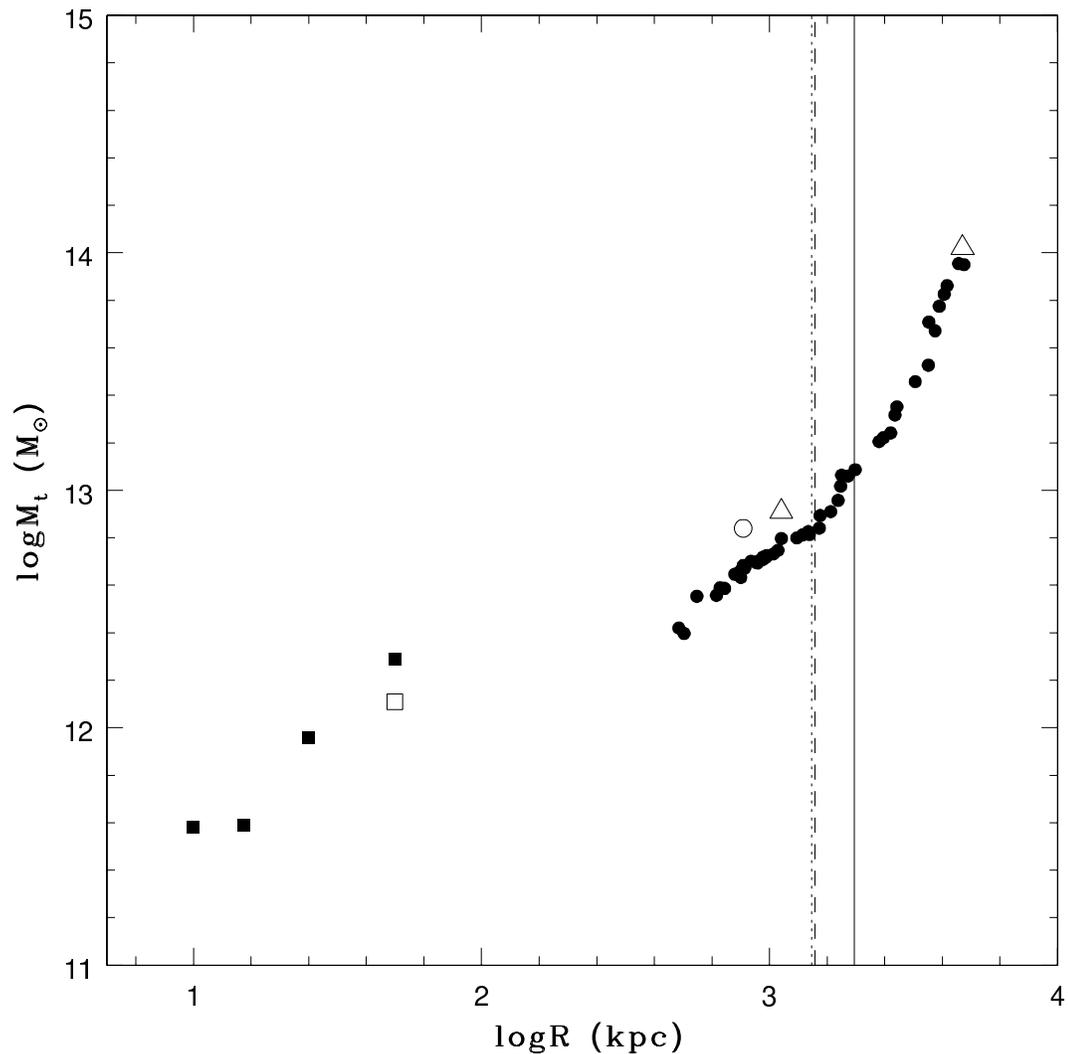}
\caption{\small{The determined total masses as a function of radius from the
  center of NGC 5128.  The symbols are the same as in
  Fig.~\ref{fig:rotamp}.  The average mass determined by
  the cumulative galaxies is $5.7\times 10^{12}$ M$_{\sun}$.  The
  radii for the dynamical and crossing timescales equal to a Hubble
  time are shown as solid and dashed lines, respectively, for all 62
  galaxies.  The dotted line is the zero velocity surface determined
  by \cite{kar06}.  The radial values plotted are the last object's radial position.}} 
\label{fig:mass}
\end{figure}

The mass was determined
separately for the 13 {\it confirmed} Centaurus group members, the M83
group members, the two independent bins of galaxies (Bin 1 and Bin 2), and for
the larger number of probable satellites
with each additional satellite galaxy added in
increasing radial projection from NGC 5128.  The adopted value of $\gamma$ was
determined independently for each list of galaxies in the same manner
as was done for the GCS in Section~\ref{mass:GCS},
with typical values ranging from $\gamma=1.5$ to $\gamma=3.2$.  The calculated
masses with projection corrections, are shown in Tables~\ref{tab:IGG}
\& ~\ref{tab:GG} and are plotted in Figure~\ref{fig:mass}.

The mass calculation for the group cannot be continued out
indefinitely far because at some radius, the fundamental assumption of
virial equilibrium breaks down.  At very large radii, the group has
not had enough time to undergo full relaxation \citep{cote97}.
The dynamical and crossing times \citep{binney87}, when set equal to a Hubble time,
occur at radii of 1967 kpc and 1400 kpc, respectively, for the entire 62
galaxies surrounding NGC 5128.  Similarly, \cite{kar06} found a
''surface of zero velocity'' around Centaurus of 1440 kpc, coinciding
with the upturn in velocity dispersion noted by the cumulative galaxy
bins seen in Fig.~\ref{fig:veldisp} and discussed in Section~\ref{kin_disc}.  Thus an
appropriate radial limit for the mass calculation is $R\simeq1.5$
Mpc.  Thus, the large masses determined for the outer 15-20 galaxies,
(includes Bin 2 and cumulative galaxy bins 42-62) are not 
valid as they are not likely virialized objects in the system.  This is
carried through to the mass-to-light ratios determined for these outer
objects in Section~\ref{masstolight}.

\subsection{Mass-to-Light Ratios}
\label{masstolight}

The mass-to-light ratios were calculated from the total mass, M$_t$,
determined in Section~\ref{mass} and a B-band luminosity
and galactic extinction for each galaxy from \cite{kar04}.

All mass-to-light ratios are listed in Tables~\ref{tab:IGG}
\&~\ref{tab:GG}.  The B-band luminosity for the galaxy calculations
include all of the galaxies within the outermost radii of each bin.  
The M83 complex, therefore, contributes to the luminosity of
Bin 1 and Bin 2 in Table~\ref{tab:IGG}, although it is not included as
part of the mass tracer population.
The B-band luminosity of NGC 5128 is $3.79\pm0.01 \times 10^{10}$
L$_{\sun}$.  This single galaxy makes up $66\%$ of the entire light of
the Centaurus group of galaxies.  The mass-to-light ratio of NGC 5128 alone 
is $52\pm22$ M$_{\sun}$/L$_{\sun}$ out to $R=45$ kpc.
    
The mass of the Centaurus group of galaxies is determined from this
study to be $(9.2\pm3.0) \times 10^{12}$ M$_{\sun}$ out to the
dynamical radius, leading to M/L$_B = 153\pm50$ M$_{\sun}$/L$_{\sun}$. 
These values are quite comparable to the most recent study of \cite{kar06}
who determined the orbital mass of the Centaurus group to be $7.5
\times 10^{12}$ M$_{\sun}$, the virial mass $8.9 \times 10^{12}$
M$_{\sun}$, and the mass determined within a zero-velocity surface
including dark energy $6.4\pm1.8 \times 10^{12}$ M$_{\sun}$.
\cite{kar06} also determined a M/L$_B = 137$ M$_{\sun}$/L$_{\sun}$.
The M/L$_B$ ratio for the
Centaurus group is larger than many other groups of galaxies
in the nearby universe (M/L$_B = 8-88$
M$_{\sun}$/L$_{\sun}$ \citep{kar05}), but mass-to-light ratios are
typically larger for groups dominated by giant ellipticals, compared to spiral
dominated groups, such as the Milky Way (M/L$_B = 28-29$
M$_{\sun}$/L$_{\sun}$), IC342 (M/L$_B = 18-30$
M$_{\sun}$/L$_{\sun}$), or M81 (M/L$_B = 19-32$ M$_{\sun}$/L$_{\sun}$) 
\citep{kar05}.  

\section{Bounding Argument}
\label{bounding}

In brief, the satellite galaxies in the Centaurus group strongly indicate a
possible connection to the halo system of its central group
member.  The {\it confirmed} members in both the Centaurus
and M83 groups, with their brightest members separated by a projected
radius of 1~Mpc, have many surrounding
galaxies with distances and/or radial velocities that do not have
confirmed group membership.  It seems a possibility that these two groups resemble the Milky
Way and M31 as bound systems.  Exploring this idea
leads to the application of a simple Newtonian bounding argument, used 
recently by \cite{brough06}, \cite{cortese04}, and others.  If the two
groups, Centaurus and M83, are part of a larger system, then the inequality
\begin{equation}
\label{eqn:bounding}
\frac{v_{r}^2 R_p}{2} \leq GM sin^2(\beta) cos(\beta) 
\end{equation}
holds.  In Equation~\ref{eqn:bounding}, $M = M83_{group} + Cen_{group} =
(9.65\pm3.01) \times 10^{12}$ M$_{\sun}$ is the total mass of the
confirmed system, $v_{r} = 25$ km s$^{-1}$ is the relative velocity between the two groups, 
$R_p = 907$ kpc is the projected distance between the
centroids of the two groups, and $\beta$ is the projection angle
between the plane of the sky and the line that joins the centers of the two groups.  The
inequality in Equation~\ref{eqn:bounding} holds for all projected angles 
between $5^o \leq \beta \leq 89^o$, using the observed radial velocities of
the Centaurus group and M83 centroids to be those of NGC 5128, $541$ km s$^{-1}$, and
NGC 5236, $516$ km s$^{-1}$, respectively.  The probability that Centaurus and M83 are
bound can then be calculated from
\begin{equation}
\label{eqn:prob}
P_{bound} = \int_{\beta1}^{\beta2} cos({\beta}) d{\beta} 
\end{equation}
\citep{girardi05}.  I find that the probability that these systems are bound is at the $91\%$ level. 

\section{Discussion and Summary}
\label{discussion}

Investigating the kinematics of the GCs in the giant
elliptical galaxy, NGC 5128, and the kinematics of 
over 60 satellite galaxies in and surrounding the Centaurus group, 
has shown interesting new results.  The rotation amplitude,
rotation axis, and velocity dispersion of both the halo of NGC 5128 and surrounding galaxies 
both have means of $\Omega R = 67\pm27$ km s$^{-1}$, $\Theta_o = 162\pm22^o$ E of
N, and $\sigma_v = 103\pm14$ km s$^{-1}$.   
The masses determined with the Tracer Mass Estimator and the
spherical component of the Jeans Equation, lead to projection
corrected masses of $(1.3\pm0.5) \times 10^{12}$ M$_{\sun}$ for NGC
5128 and  $(9.2\pm3.0) \times 10^{12}$ M$_{\sun}$ for the entire
Centaurus group.  Mass-to-light ratios in the B-magnitude were subsequently determined to
be $52\pm22$ M$_{\sun}$/L$_{\sun}$ for NGC 5128 and $153\pm50$
M$_{\sun}$/L$_{\sun}$ for the Centaurus group.

The similar kinematics between the halo of the brightest group member
with its surrounding satellites suggests that the halo of 
NGC 5128 could therefore be thought of as the innermost
component of the entire group that surrounds it. 

The Centaurus group is likely bound, although its outer regions are
unlikely to be virialized beyond $R\simeq1.5$ Mpc.  The global dynamics of the group that
we are seeing, could therefore represent the
initial internal shear of the pregalactic filamentary structure from which it
condensed.   NGC 5128 could have formed early on as a galaxy that was already
centrally dominant, and now continues to build up by a
series of minor mergers and accretions, leading to an averaging
of the kinematics from the surrounding satellite galaxies that we now
measure in the GCs in the NGC 5128 halo.  A single
or small number of major merger events would not necessarily produce these
kinematic similarities.  This view is consistent with the recent work
on the halo stars in NGC 5128 which have a mean age of $8^{+3}_{-3.5}$
Gyr \citep{rejkuba05}, and the age distribution of the GCs which shows
that a high fraction of both metal-rich and metal-poor GCs are old
(M. Beasley et al. 2006, in preparation).  It would be interesting to extend this type of
study to other nearby groups with centrally dominant giants, although
large samples of GCs and satellite galaxies with measured radial
velocities are required in each case.

\acknowledgments
K.A.W. would like to thank Willam E. Harris and James Wadsley 
from McMaster University for their helpful
suggestions and many insightful discussions, as well as Mike Beasley 
for transmitting cluster velocity data in
advance of publication.  K.A.W. would also like
to thank the referee who provided very thorough comments on this
paper.  This research was funded
by NSERC through grants to W.E.H.




\clearpage
		   		
\begin{deluxetable}{rrrrrrrllll}
\tablecaption{Kinematic and Dynamic Solutions for the Globular Cluster System of NGC 5128 \label{tab:GC}}
\tablewidth{0pt}
\rotate
\tablehead{
\colhead{Radial Bin} & {R$_{mean}$} &  {R$_{out}$} & \colhead{N} & \colhead{$\Omega
  R$} & \colhead{$\Theta_o$} & \colhead{$\sigma_v$} & \colhead{M$_{corr}$} & \colhead{M$_{p}$} & \colhead{M$_{r}$} & \colhead{M$_{t}$}\\
\colhead{(kpc)} & \colhead{(kpc)} & \colhead{(kpc)}& \colhead{ } & \colhead{(km
  s$^{-1}$)} & \colhead{($^o$ E of N)} & \colhead{(km s$^{-1}$)}&
\colhead{ } & \colhead{($\times$ 10$^{10}$ M$_{\sun}$)} 
& \colhead{($\times$ 10$^{10}$ M$_{\sun}$)} & \colhead{($\times$ 10$^{10}$ M$_{\sun}$)} \\
}

\startdata 
0-50  & 12.9 & 48.8 & 343 & 40$\pm$10 & 189$\pm$12 & 123$\pm$6  & 1.0 & 125.8$\pm$46.5 & 3.1$\pm$1.3  & 128.9$\pm$46.5\\
0-5   & 3.64 & 4.98 & 55  & 24$\pm$21 & 334$\pm$59 & 120$\pm$13 & -   &   -            &  -           &   -           \\
5-10  & 7.65 & 9.96 & 124 & 43$\pm$15 & 195$\pm$20 & 112$\pm$8  & 2.3 &  37.4$\pm$6.4  & 0.4$\pm$0.3  &  37.8$\pm$6.4\\
10-15 & 12.4 & 14.9 & 68  & 83$\pm$25 & 195$\pm$12 & 105$\pm$11 & 1.6 &  36.5$\pm$9.3  & 2.4$\pm$1.5  &  38.9$\pm$9.4\\
15-25 & 19.0 & 24.3 & 57  & 35$\pm$26 & 183$\pm$34 & 147$\pm$16 & 1.3 &  89.5$\pm$27.4 & 0.7$\pm$1.0  &  90.2$\pm$27.4\\
25-50 & 34.7 & 48.8 & 39  & 96$\pm$45 & 169$\pm$17 & 148$\pm$22 & 1.0 & 184.7$\pm$84.5 & 10.4$\pm$9.8 & 195.1$\pm$85.1\\
\enddata

\end{deluxetable}
\clearpage
\thispagestyle{empty}

\begin{deluxetable}{rrrrrrlllllcl}
\tablecaption{Kinematic and Dynamic Solutions for the Centaurus Group,
  M83 Group, and Radially Binned Galaxies Surrounding NGC 5128\label{tab:IGG}}
\tablewidth{0pt}
\tabletypesize{\small}
\rotate
\tablehead{
\colhead{Bin} & \colhead{R$_{mean}$} & \colhead{R$_{out}$} &\colhead{$\Omega R$} &
\colhead{$\Theta_o$} & \colhead{$\sigma_v$} & \colhead{$\gamma$} &
\colhead{M$_{p}$} & \colhead{M$_{r}$} &
\colhead{M$_{t}$} & \colhead{L$_B$} & \colhead{M/L$_B$}\\
\colhead{ } & \colhead{(kpc)} & \colhead{(kpc)} & \colhead{(km s$^{-1}$)} &
\colhead{($^o$ E of N)} & \colhead{(km s$^{-1}$)} & \colhead{ } 
& \colhead{($\times$ 10$^{11}$ M$_{\sun}$)} 
& \colhead{($\times$ 10$^{11}$ M$_{\sun}$)} 
& \colhead{($\times$ 10$^{11}$ M$_{\sun}$)} 
& \colhead{($\times$ 10$^{10}$ L$_{\sun}$)} 
& \colhead{(M$_{\sun}$/L$_{\sun}$)} \\ 
}
\startdata
M83&  119.1\tablenotemark{a} & 237.1\tablenotemark{a} & 25$\pm$27 & 299$\pm$88 &  53$\pm$16 &1.79$\pm$0.67 & 4.7$\pm$2.7   &  0.3$\pm$0.7   & 5.0$\pm$2.8     & 4.0  &   12$\pm$6   \\
Cen&  358.5 & 810.1 &125$\pm$50 & 159$\pm$23 & 115$\pm$25 &2.23$\pm$0.31 & 62.1$\pm$18.7 &  29.5$\pm$23.6 & 91.5$\pm$30.1  & 6.0  & 153$\pm$50  \\
Bin 1\tablenotemark{b} &539.5  & 1099.1  & 67$\pm$30 & 156$\pm$24 &  98$\pm$15 &2.01$\pm$0.10 & 69.5$\pm$6.2  &  11.6$\pm$10.3 & 81.1$\pm$12.0   & 10.0  & 81$\pm$12   \\ 
Bin 2\tablenotemark{c} & 2594.2 & 4727.75 & 36$\pm$58 & 108$\pm$50 & 179$\pm$26 &3.17$\pm$0.40 & 1052$\pm$363  &  14.1$\pm$45.7 & 1066$\pm$366   & 13.9  & 765$\pm$263 \\  
\enddata

\tablenotetext{a}{The projected radius from NGC 5236.}
\tablenotetext{b}{Bin 1 includes the 27 galaxies nearest NGC 5128 with
  radial velocities, excluding the M83 group.}
\tablenotetext{c}{Bin 2 includes the 28$^{th}$ to 53$^{rd}$ nearest galaxies to NGC 5128 with radial velocities, excluding the M83 group.}
\end{deluxetable}
\clearpage

\begin{deluxetable}{rrrrrlllllcl}
\tablecaption{Kinematic and Dynamic Solutions for the Cumulatively Binned Galaxies Surrounding NGC 5128 \label{tab:GG}}
\tablewidth{0pt}
\tabletypesize{\tiny}
\tablehead{
\colhead{N} & \colhead{R$_{out}$} &\colhead{$\Omega R$} &
\colhead{$\Theta_o$} & \colhead{$\sigma_v$} & \colhead{$\gamma$} &
\colhead{M$_{p}$} & \colhead{M$_{r}$} &
\colhead{M$_{t}$} & \colhead{L$_B$} & \colhead{M/L$_B$}\\
\colhead{ } & \colhead{(kpc)} & \colhead{(km s$^{-1}$)} &
\colhead{($^o$ E of N)} & \colhead{(km s$^{-1}$)} & \colhead{ } 
& \colhead{($\times$ 10$^{11}$ M$_{\sun}$)} 
& \colhead{($\times$ 10$^{11}$ M$_{\sun}$)} 
& \colhead{($\times$ 10$^{11}$ M$_{\sun}$)}  
& \colhead{($\times$ 10$^{10}$ L$_{\sun}$)} 
& \colhead{(M$_{\sun}$/L$_{\sun}$)} \\
}

\startdata
14 &  484.5 & 74$\pm$38 & 157$\pm$32 &  90$\pm$20 &1.56$\pm$0.11 & 20.1$\pm$1.6  &  6.2$\pm$6.3   & 26.3$\pm$6.5    & 4.1  &  64$\pm$16  \\
17 &  653.9 & 76$\pm$35 & 155$\pm$31 &  92$\pm$17 &1.52$\pm$0.08 & 27.3$\pm$1.6  &  8.8$\pm$8.1   & 36.1$\pm$8.3    & 5.4  &  66$\pm$15  \\
20\tablenotemark{d} &  757.0 & 72$\pm$33 & 152$\pm$28 &  93$\pm$16 &1.83$\pm$0.02 & 35.2$\pm$1.0  &  9.1$\pm$8.4   & 44.3$\pm$8.4    & 5.5  &  79$\pm$15  \\
23\tablenotemark{d} &  795.5 & 77$\pm$31 & 153$\pm$22 &  90$\pm$15 &1.83$\pm$0.02 & 35.0$\pm$1.0  &  11.0$\pm$8.8  & 46.0$\pm$8.9    & 5.7  &  80$\pm$15  \\
26\tablenotemark{d} &  862.4 & 88$\pm$29 & 154$\pm$17 &  87$\pm$13 &1.53$\pm$0.29 & 34.8$\pm$7.0  &  15.6$\pm$10.3 & 50.3$\pm$12.4   & 5.7  &  87$\pm$21  \\
29\tablenotemark{d} &  948.2 & 86$\pm$28 & 156$\pm$16 &  84$\pm$12 &1.75$\pm$0.30 & 35.9$\pm$8.1  &  16.3$\pm$10.6 & 52.3$\pm$13.3   & 9.7  &  54$\pm$13  \\
32\tablenotemark{d} &  976.5 & 68$\pm$24 & 143$\pm$20 &  87$\pm$12 &1.75$\pm$0.30 & 42.3$\pm$9.5  &  10.5$\pm$7.4  & 52.8$\pm$12.1   & 9.7  &  54$\pm$12  \\
35\tablenotemark{d} & 1069.5 & 64$\pm$25 & 154$\pm$20 &  88$\pm$12 &1.68$\pm$0.22 & 45.7$\pm$7.4  &  10.2$\pm$8.0  & 55.9$\pm$10.9   & 9.7  &  57$\pm$11  \\
38\tablenotemark{d} & 1300.2 & 61$\pm$25 & 159$\pm$19 &  90$\pm$12 &1.68$\pm$0.22 & 53.6$\pm$8.6  &  11.3$\pm$9.2  & 64.8$\pm$12.6   & 9.8  &  66$\pm$12  \\
41\tablenotemark{d} & 1488.1 & 61$\pm$23 & 164$\pm$18 &  88$\pm$12 &1.68$\pm$0.22 & 56.4$\pm$9.1  &  12.9$\pm$9.7  & 69.3$\pm$13.3   & 9.9  &  69$\pm$13  \\
44\tablenotemark{d} & 1728.8 & 69$\pm$23 & 175$\pm$15 &  91$\pm$11 &1.94$\pm$0.29 & 71.5$\pm$18.1 &  19.2$\pm$12.8 & 90.6$\pm$22.2   & 13.1 &  69$\pm$16  \\
47\tablenotemark{d} & 1874.2 & 71$\pm$24 & 185$\pm$15 &  97$\pm$12 &1.94$\pm$0.29 & 92.8$\pm$23.6 &  22.0$\pm$14.9 & 114.8$\pm$27.9  & 13.2 &  87$\pm$21  \\
50\tablenotemark{d} & 2486.4 & 69$\pm$23 & 188$\pm$15 &  99$\pm$11 &2.25$\pm$0.32 & 139.0$\pm$47.6&  27.6$\pm$18.4 & 166.5$\pm$50.99 & 13.7 &  121$\pm$37  \\
53\tablenotemark{d} & 2767.2 & 47$\pm$24 & 191$\pm$24 & 104$\pm$11 &2.25$\pm$0.32 & 210.5$\pm$72.5&  14.2$\pm$14.5 & 224.7$\pm$73.93 & 13.8 &  162$\pm$53  \\
56\tablenotemark{d} & 3572.3 & 49$\pm$26 & 177$\pm$24 & 117$\pm$11 &2.57$\pm$0.28 & 490.6$\pm$218.6& 20.0$\pm$21.2 & 510.6$\pm$219.7 & 13.8 &  368$\pm$158  \\
59\tablenotemark{d} & 4041.1 & 41$\pm$27 & 158$\pm$31 & 117$\pm$11 &2.47$\pm$0.28 & 653.5$\pm$251.4& 15.8$\pm$20.8 & 669.3$\pm$252.3 & 13.9 &  481$\pm$181  \\
62\tablenotemark{d} & 4727.8 & 44$\pm$27 & 137$\pm$32 & 137$\pm$12 &2.47$\pm$0.28 & 869.8$\pm$337.8& 21.3$\pm$26.2 & 891.1$\pm$338.8 & 13.9 &  640$\pm$243  \\
	       			
\enddata       

\tablenotetext{d}{Includes confirmed galaxies from the M83 group.}

\end{deluxetable}

																	      
\clearpage

\end{document}